\documentclass[10pt,a4paper]{article}

\usepackage{graphicx}
\usepackage[latin1]{inputenc}

\title{
\textbf{
Dielectric study of the glass transition of PET/PEN blends.
}
}

\author{
J.~Sellarès\thanks{e-mail:{\tt jordi.sellares@upc.edu}}, J.A.~Diego, J.C.~Cañadas, M.~Mudarra and J.~Belana\\  
\em\small{Departament de Física i Enginyeria Nuclear, Universitat Politècnica de Catalunya}\\
\em\small{Campus de Terrassa, c. Colom~1, E-08222 Terrassa, Spain.}\\
P.~Colomer, F.~Román and Y.~Calventus\\
\em\small{Departament de Màquines i Motors Tèrmics, Universitat Politècnica de Catalunya}\\
\em\small{Campus de Terrassa, c. Colom~11, E-08222 Terrassa, Spain.}\\
}

\date{}

\begin{document}

\maketitle

\begin{abstract}

An analysis of the glass transition of four materials with similar chemical structures is performed: PET, PEN and two PET/PEN blends (90/10 and 70/30 w/w). During the melt processing of the blends transesterification reactions yield block and random PET/PEN copolymers that act as compatibilizers. The blends obtained in this way have been characterized by $^1$H--NMR and DSC. A degree of randomness of $0.38$ and $0.26$ has been found for the 90/10 and 70/30 copolymers. It is shown by DSC that this copolimerization is enough to compatibilize the blends. The $\alpha$ relaxation, the dielectric manifestation of the glass transition, has been studied by thermally stimulated depolarization currents (TSDC). The relaxation has been analyzed into its elementary modes by means of a relaxation map analysis. The activation energies of the modes of the glass transition do not change significantly between the four materials: in all cases the modes with a larger contribution have around $3$~eV and modes with less than $1$~eV are not detected. The change in the pre--exponential factor accounts entirely for the relaxation time change from material to material, that is larger as the PEN content increases. The compensation law is fulfilled and compensation plots converge for high--frequency modes. The polarizability decreases as the PEN content increases due to the increased stiffness of the polymer backbone. An analysys of the cooperativity shows that the central modes of the distribution are the most cooperative while high--frequency modes tend to behave more as Arrhenius. The low--frequency modes are difficult to study due to the asymmetry of the distribution of relaxation times. PEN turns out to be the less cooperative material. It is demonstrated how the parameters obtained from the dielectric study are able to reproduce calorimetric data from DSC scans and are, therefore, a valid description of the glass transition.

\end{abstract}


\newpage

\section{Introduction}
\label{intro}

\begin{figure}
\begin{center}
\includegraphics[width=10cm,clip]{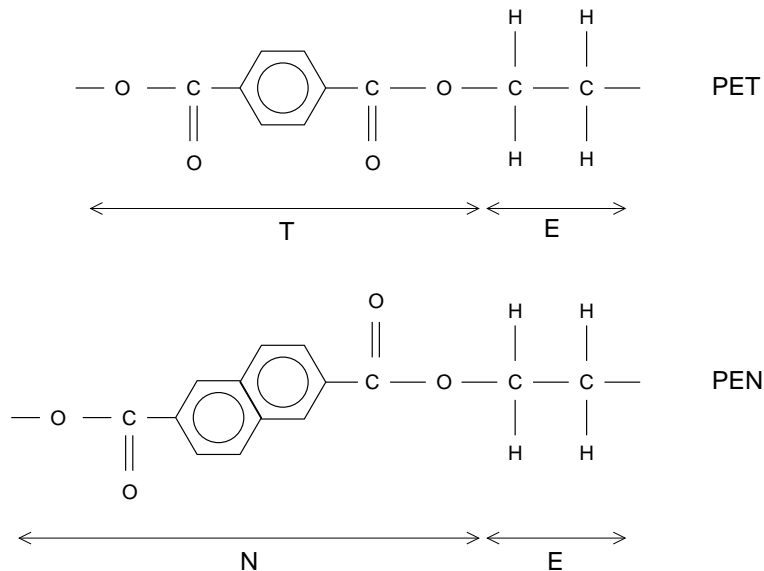}
\end{center}
\caption{Repeating units of PET and PEN.}
\label{petpen}
\end{figure}

Poly (ethylene terephthalate) (PET) and  poly (ethylene naphthalate) (PEN) are two commercially important polyesters that have a similar structure, as it can be seen in figure~\ref{petpen}.
Indeed, the only difference between them is that the single aromatic ring in PET is replaced by a double one in PEN. For this reason, PEN has improved thermal and mechanical properties over PET even though both materials share some desirable properties such as low permeability to gases, good impact strength and resistance to chemical attack. In spite of this fact, commercial applications of PEN are limited because it is more expensive than PET.

PET/PEN copolymers are interesting because they have properties that are intermediate to both materials at a lower cost than pure PEN. PET and PEN are immiscible \cite{porter} but when transesterification takes place the copolymers that are formed act as compatibilizers and, therefore, miscible blends can be obtained \cite{shi}. In a transesferification reaction a PET repeating unit and a PEN repeating unit interchange their positions  
\[ \textrm{T-E-T} + \textrm{N-E-N} \rightleftharpoons \textrm{T-E-N} + \textrm{N-E-T} \]
where E stands for ethylene group and the units are part of a polymer chain. At the beginning, only block copolymers are formed but after many interchange reactions random copolymers are also obtained \cite{eguiazabal}. The statistical parameter that quantifies this effect is the degree of randomness, which is close to $0$ for block copolymers and equal to $1$ for random copolymers. Transesterification reactions take place, for example, during melt processing of a blend. 

It has been shown that after a certain transesterification level has been reached, the properties of the blend depend only on the blend composition rather than on the degree of randomness \cite{shi}. Therefore, it is not necessary to obtain completely random copolymers to obtain blends with stable properties, intermediate to those of PET and PEN.

As any other polymer or copolymer, PET/PEN copolymers have a glass transition. The glass transition is a phenomenon related to the structural relaxation of the material. Above the glass transition temperature ($T_g$), the material is always at structural equilibrium and its free volume depends only on the temperature ($T$). If the material is cooled, at first the free volume is able to diminish accordingly to the temperature but once the glass transition temperature is reached the system falls out of structural equilibrium and the free volume no longer corresponds to the current temperature. The material is said to be in the glassy state and, aside from the temperature, its state is characterized by the so--called fictive temperature ($T_f$). The fictive temperature of the material at a given state is the temperature at which the material at structural equilibrium would have the same free volume as in the current state. In fact, when the material is at a constant temperature lower that $T_g$ the system tends towards structural equilibrium so the fictive temperature tends to attain the same value as the temperature. This process is known as physical aging.

Immiscible blends of two components present two glass transitions but once transesterification has taken place, the blend of PET and PEN becomes miscible and a single glass transition is observed. Moreover, once a certain value of the degree of randomness is attained, the glass transition temperature ceases to depend on this magnitude \cite{shi}.

The glass transition of PET/PEN copolymers has been throughoutly studied through dielectric properties, using mainly dielectric loss spectroscopy. It has been found that the dielectric strength and the width of the relaxation time distribution are sensible to the processing parameters but molecular coupling does not change significantly with the blend composition or the degree of transesterification \cite{copos1}. The $\alpha$ relaxation, the dielectric manifestation of the glass transition \cite{peta}, depends strongly on the thermal history \cite{copos2} because crystallization and physical aging affects the amorphous phase that is where this phenomenon takes place \cite{segona}. The relative content of PET and PEN also affects the $\alpha$ relaxation. Particularly, there is a noticeable reduction in the dielectric strength due to a decrease in the mobility of local segments as the content of PEN increases \cite{copos5}. With regards to other relaxations, a subglass relaxation $\beta^*$ can be observed in PEN and copolymers but not in PET \cite{copos2, copos4}. 

In a previous work \cite{petrma}, we demonstrated that the glass transition could be studied by dielectric means using the thermally stimulated depolarization currents (TSDC) technique to perform a relaxation map analysis (RMA) \cite{laca}. The poling schemes used in the RMA must excite only elementary modes of the relaxation. To this end, non--isothermal window poling (NIW) or window poling (WP) are appropriate choices. Within a RMA, a depolarization current is obtained for each elementary mode and then the parameters of each mode are obtained fitting the data to a relaxation model such as Arrhenius or Tool--Narayanaswamy--Moynihan (TNM). The later is able to model the structural state of the system and therefore it is suitable to model memory effects or physical aging. This approach has the advantage that a distributed process such as the glass transition can be analyzed as elementary modes.

The goal of this work is to understand in which way the glass transition of PET/PEN blends/copolymers differ. Since both polymers are very similar we expect that some parameters of the glass transition will remain unchanged while some others will change gradually between the PET and the PEN values in terms of the PET/PEN ratio. These parameters will be obtained following the aforementioned TSDC--RMA approach. Throughout this work we will take advantage of the distinctive capability of this approach to study individual modes on its own. Since we will be studying the glass transition using dielectric techniques, we will also check that the parameters obtained reproduce data from calorimetric experiments and, therefore, describe the glass transition satisfactorily.

\section{Experimental}

\label{exp}

\subsection{Preparation}

A commercial PET, with a number--average molecular weight ($M_n$) of $2.0 \times 10^4$~\cite{pes} and a commercial PEN, Kaladex, supplied by Goodfellow, with a density of $1.36$~g$\cdot$cm$^{-3}$ and nominal granule size of 5~mm were used in this work.

Previous to the preparation of the blends, the polymers were conditioned for $12$~h at $150$~$^\circ$C in a vacuum oven, in order to eliminate the absorbed water and minimize possible hydrolysis during the melting process.

PET/PEN blends, from 90/10 and 70/30 (w/w) mixtures, were prepared by melt extrusion using a twin screw extruder Haake Rheocorder with a screw speed of 50~rpm. Samples of 60~g, approximately, were melted at 290~$^\circ$C for 5 minutes. A nitrogen purge was used during processing to prevent oxidative degradation. The extrudate passed through a cooling ice--water bath and was vacuum dried at 70~$^\circ$C in order to eliminate absorbed water. 

PET and PEN pure polymers were processed in the extruder at the same conditions as the blends so finally four materials have been studied: PET, 90/10, 70/30 and PEN.

The samples obtained from the extruder were melted at 290~$^\circ$C. After 5 minutes at this temperature they were molded and quenched to room temperature by two parallel plates refrigerated by water. A thickness of 0.5~mm was obtained for all these sheets.

The sheets obtained in this way had low cristallinity, as indicated by the transparency of the sheets and the lack of milky patches. The cristallinity degree was estimated from differential scan calorimetry (DSC) measurements and found to be less than 10\% in all cases. Also, it should be noted that a small degree of crystallinity poses no problem since the $\alpha$ relaxation of the interspherulitic amorphous phase is analogous to the $\alpha$ relaxation of the amorphous material and is easy to distinguish from the $\alpha$ relaxation of the amorphous interlamellar phase ($\alpha_c$) \cite{segona}.

\subsection{Characterization of the blends}

The composition determination and the sequence distribution analysis of the copolyesters were conducted by $^1$H--NMR spectra using a Bruker AMX--300 spectrometer at 25~$^\circ$C operating at 300 MHz. The samples (20~mg) were dissolved in 1~mL of a mixture of deuterated chloroform (CDCl$_3$) / trifluoroacetic acid (TFA) (8/1 v/v) and spectra were internally referenced to tetramethylsilane (TMS)~\cite{japos}. Sixty--four scans were required with 32K data points and a relaxation delay of 20~s. Integration of the overlapping signals was made by lorentzian deconvolution of the spectra using the Bruker 1D WIN--NMR computer software.

DSC scans were obtained with a Mettler DSC--20 calorimeter controlled by a TC11 processor. Experiments were performed on $20$~mg samples sealed on aluminum pans. DSC scans with a heating rate of $2.5$~$^\circ$C/min were performed on these samples.

\subsection{Measurements}

The experimental setup for TSDC consists of a custom build forced air oven controlled by a Eurotherm--2416 PID temperature programmer and a Knürr-Heinzinger (6~kV) potential source. The samples are positively biased connecting one electrode to the source while the other one is grounded. The temperature is measured to an accuracy of $0.1$~K by a thermocouple sensor located close to the samples. A Keithley--6514 electrometer was employed for the current intensity measurements.

The sheets were cut into appropriate samples for measurement by TSDC. Circular aluminium electrodes with $1$~cm diameter were vacuum deposited on each side of the samples. 

\begin{figure}
\begin{center}
\includegraphics[width=10cm,clip]{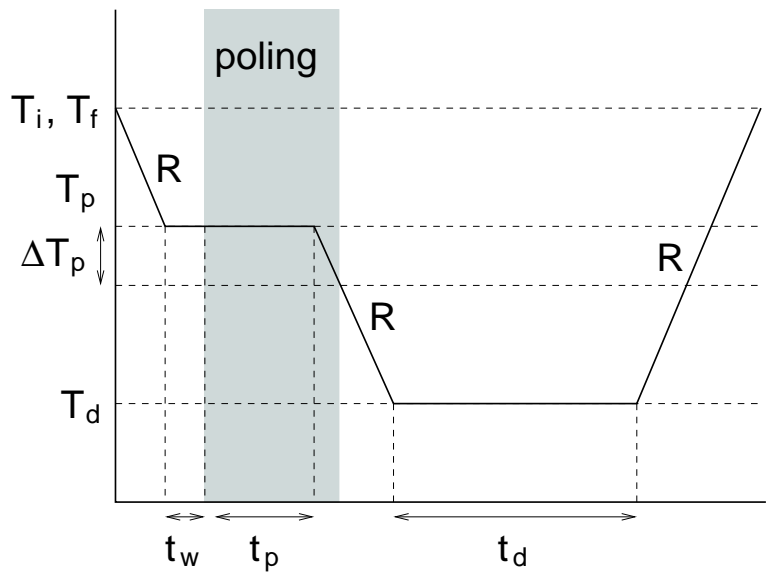}
\end{center}
\caption{Outline of a TSDC experiment.}
\label{tsdc}
\end{figure}

An outline of a conventional TSDC experiment can be seen in figure~\ref{tsdc}. 
At the beginning of the experiment the sample is at an initial temperature $T_i$. It is brought to a poling temperature $T_p$ and after a short time $t_w$ it is poled for a poling time $t_p$. Then, the sample is cooled down to a diposit temperature $T_d$. During this cooling ramp, the poling field is switched off at a temperature $\Delta T_p$ below $T_p$. After a diposit time $t_d$ it is heated up to a final temperature $T_f$ while the depolarization current is recorded. The depolarization current represented in front of $T$ is also called a spectrum or a thermogram. 

A RMA was performed on a sample of each material using two different poling schemes. In the first scheme, non--isothermal window poling (NIW), there is no isothermal poling stage ($t_w=t_p=0$) so the poling field is applied entirely during a cooling ramp. It is switched on at a polarization temperature $T_p$ and removed when the sample reaches a temperature $T_p - \Delta T_p$. If $\Delta T_p$ is small enough, the depolarization current behaves almost as coming form an elementary mode. NIW poling has the advantage that all the measurements can be performed with the same thermal history even if they have different $T_p$. The value of $\Delta T_p$ employed ($2$~$^\circ$C) is the smallest one that still yields a depolarization current that can be recorded with an acceptable resolution with our equipment. The parameters employed in our experiments are enclosed in table~\ref{parameters-niw}.
\begin{table}

\caption{Parameters used in the NIW RMA.\label{parameters-niw}}

\begin{center}

\begin{tabular}{c|c|c|c|c|c|c}
$V$      & $R$                  & $t_w $  & $t_p$   & $\Delta T_p$  & $T_d$          & $t_d$ \\ \hline
$1500$~V & $2.5$~$^\circ$C/min  & $0$     & $0$     & $2$~$^\circ$C & $20$~$^\circ$C & $10$~min \\
\end{tabular}

\medskip

\begin{tabular}{c|c|c}
PET/PEN & $T_i$, $T_f$    & $\{T_p\}$ \\ \hline
PET     & $90$~$^\circ$C  & $78$~$^\circ$C to $52$~$^\circ$C in $2$~$^\circ$C steps \\
90/10   & $96$~$^\circ$C  & $82$~$^\circ$C to $58$~$^\circ$C in $2$~$^\circ$C steps \\
70/30   & $102$~$^\circ$C & $90$~$^\circ$C to $64$~$^\circ$C in $2$~$^\circ$C steps \\
PEN     & $132$~$^\circ$C & $122$~$^\circ$C to $94$~$^\circ$C in $2$~$^\circ$C steps \\
\end{tabular}

\end{center}

\end{table}

As saturation is usually not attained using NIW poling, another RMA was performed for each material using the window poling scheme (WP), to obtain data about polarizability. In this scheme poling is performed isothermally at $T_p$ and the electric field is switched off at the end of the isothermal poling stage, just when the sample cooling begins ($\Delta T_p = 0$). As most poling temperatures are just below $T_g$, the sample will undergo physical aging during the poling stage. A moderate poling time ($5$~min) was employed to limit physical aging during the poling stage and, therefore, ensure that the fictive temperature and, consequently, the relaxation time do not change too much during the isothermal stage. This minimizes the effect of the different thermal history of the samples in each measurement. The parameters employed in our experiments are enclosed in table~\ref{parameters-wp}.
\begin{table}

\caption{Parameters used in the WP RMA.\label{parameters-wp}}

\begin{center}

\begin{tabular}{c|c|c|c|c|c|c}
$V$      & $R$                  & $t_w $  & $t_p$   & $\Delta T_p$  & $T_d$          & $t_d$ \\ \hline
$1500$~V & $2.5$~$^\circ$C/min  & $1$~min & $5$~min & $0$           & $20$~$^\circ$C & $10$~min \\
\end{tabular}

\medskip

\begin{tabular}{c|c|c}
PET/PEN & $T_i$, $T_f$    & $\{T_p\}$ \\ \hline
PET     & $90$~$^\circ$C  & $78$~$^\circ$C to $48$~$^\circ$C in $2$~$^\circ$C steps \\
90/10   & $96$~$^\circ$C  & $82$~$^\circ$C to $54$~$^\circ$C in $2$~$^\circ$C steps \\
70/30   & $102$~$^\circ$C & $90$~$^\circ$C to $60$~$^\circ$C in $2$~$^\circ$C steps \\
PEN     & $132$~$^\circ$C & $122$~$^\circ$C to $90$~$^\circ$C in $2$~$^\circ$C steps \\
\end{tabular}

\end{center}

\end{table}

In a previous work \cite{petrma} it was found that NIW with $\Delta T_p = 2$~$^\circ$C and WP with $t_p = 5$~min gave similar values of the relaxation time of each mode and of their relative weight, in the case of PET. The agreement between these results is an indication that, on the one hand, $\Delta T_p$ is small enough to obtain spectra from mainly one elementary mode and, on the other hand, $t_p$ is small enough so that there is no significant change on the relaxation time due to physical aging during the isothermal stage.

To check the validity of the modeling of the glass transition, DSC scans were performed with different cooling rates. These scans were also made with the Mettler DSC--20 calorimeter with samples prepared in the same way as for the characterization of the blends. In these experiments each sample is heated at the same initial temperature as in the TSDC experiments. Then it is cooled at a cooling rate $R_c$ of $1.25$, $2.5$ or $5$~$^\circ$C/min until it arrives to $20$~$^\circ$C. The sample is kept for $10$~min at this temperature and then it is heated at a rate of $2.5$~$^\circ$C/min.

\section{Results and discussion}
\label{resdis}

\subsection{Characterization of the blends results}

In the $^1$H--NMR spectra, the peak of the ethylene links N--E--T (between naphthalate and terephthalate) appears at approximately $4.84$~ppm, while those for N--E--N links (between two naphthalate units) and T--E--T links (between two terephthalate units) appear at approximately $4.89$~ppm and $4.79$~ppm, respectively. The evaluation of the extent of transesterification and the resulting sequence distribution were made from the areas under these peaks~\cite{shi2, jun}. Table~\ref{randomness}
\begin{table}

\caption{Composition and sequence distribution of PET/PEN blends.}
\label{randomness}
\footnotesize
\begin{center}
\begin{tabular}{|c|cc|ccc|cc|c|}\hline
Blend & \multicolumn{2}{c|}{Molar}       & \multicolumn{3}{c|}{Dyads}     & \multicolumn{2}{c|}{Number average}  & Random- \\
w/w   & \multicolumn{2}{c|}{composition} & \multicolumn{3}{c|}{(mol--\%)} & \multicolumn{2}{c|}{(Sequence length)}& ness \\
      &  $x_{\rm ET}$ & $x_{\rm EN}$     & $P_{\rm TET}$ & $P_{\rm TEN}$  & $P_{\rm NEN}$   & $n_{\rm ET}$ & $\rm n_{EN}$ & $\chi$     \\ \hline
PET/PEN & 92.2 & 7.8 & 88.4   & 6.0    & 5.6    & 30.5    & 2.9   & 0.38   \\
90/10   &      &     & (85.0) & (14.4) &  (0.6) &  (12.8) & (1.1) & (1.00) \\ \hline
PET/PEN & 74.5 & 25.5 & 68.4   & 10.0   & 21.5  & 14.7  & 5.3   & 0.26  \\
70/30   &      &      & (55.5) & (38.0) & (6.5) & (3.9) & (1.3) & (1.00) \\ \hline
\end{tabular}
\end{center}

\end{table}
shows the obtained results of the composition and of the sequence distribution of the blends.

Molar composition of the blends, $x_{\rm ET}$ and $x_{\rm EN}$, was determined from the aromatic proton resonances observed in $^1$H--NMR spectra. The mole fraction ($P$) of dyads of T--E--N, T--E--T and N--E--N units, $P_{\rm TEN}$, $P_{\rm TET}$ and $P_{\rm NEN}$, are obtained from the intensities or areas of these three peaks ($A_{\rm TET}$, $A_{\rm TEN}$ and $A_{\rm NEN}$, respectively)
\begin{eqnarray}
P_{\rm TET} & = & A_{\rm TET} / A \\
P_{\rm NEN} & = & A_{\rm NEN} / A \\
P_{\rm TEN} & = & A_{\rm TEN} / A                      
\end{eqnarray}
where $A = A_{\rm TET} + A_{\rm TEN} + A_{\rm NEN}$.      
 
The degree of randomness ($\chi$) is determined by the equation:
\begin{equation}
\chi = P_{\rm NT} + P_{\rm TN} = \frac{P_{\rm NET}}{2 P_{\rm t}} + \frac{P_{NET}}{2 P_{\rm n}}
\end{equation}   
where $P_{\rm n} = \frac{P_{\rm NET}}{2} + P_{\rm NEN}$  is the mole fraction of naphthalate units, $P_{\rm t} = \frac{P_{\rm NET}}{2} + P_{\rm TET}$ is the mole fraccion of terephthalate units, $P_{\rm NT}$ is the probability of finding an N unit next to a T unit and $P_{\rm TN}$ is the probability of finding a T unit next to an N unit. From probability theory $P_{\rm t} = 1 - P_{\rm n}$. Therefore, 
\begin{equation}
\chi = \frac{P_{\rm NET}}{2 P_{\rm n} (1 - P_{\rm n})}
\end{equation}
 
According Shi and Jabarin \cite{shi, shi2}, if $\chi = 1$, the N and T units take a totally random distribution, that is, the copolymer formed is a random copolymer. If $\chi < 1$, the N and T units tend to cluster in blocks of each one of the units, that is, the sequence length is long and the copolymer formed is a block copolymer. If $\chi > 1$, the length of the sequences becomes shorter and the copolymer tends to form an alternating copolymer. If $\chi = 2$, an alternating copolymer is formed. If $\chi = 0$, the system is a mixture of homopolymers. The randomness degree of the blends used in this work are $0.38$ and $0.26$, respectively. This fact indicates the presence of a block copolymer in  both samples as consequence of the transesterification reaction. The randomness degree decreases with the increasing of PEN content.
 
The number average sequence length of N and T units in the polymer are given by
\begin{equation} 
n_{\rm EN} = \frac{2 P_{\rm n}}{P_{\rm NET}} = \frac{1}{P_{\rm NT}}
\end{equation}
and
\begin{equation} 
n_{\rm ET} = \frac{2 P_{\rm t}}{P_{\rm NET}} = \frac{1}{P_{\rm TN}}
\end{equation}
 
In table~\ref{randomness} it can be seen that a  very low value of $2.9$ is obtained for the 90/10 blend. As the PEN content is larger in the blend, the number average sequence length of EN increases ($2.9$ for the 90/10 sample and $5.3$ for the 70/30 sample) and the value of ET decreases ($30.5$ for the 90/10 sample and $14.7$ for the 70/30 sample). In this table we also present for comparison, in brackets, the theoretical values calculated on the basis of a random comonomer distribution using the copolyester composition data given in this table.

\begin{figure}
\begin{center}
\includegraphics[width=10cm,clip]{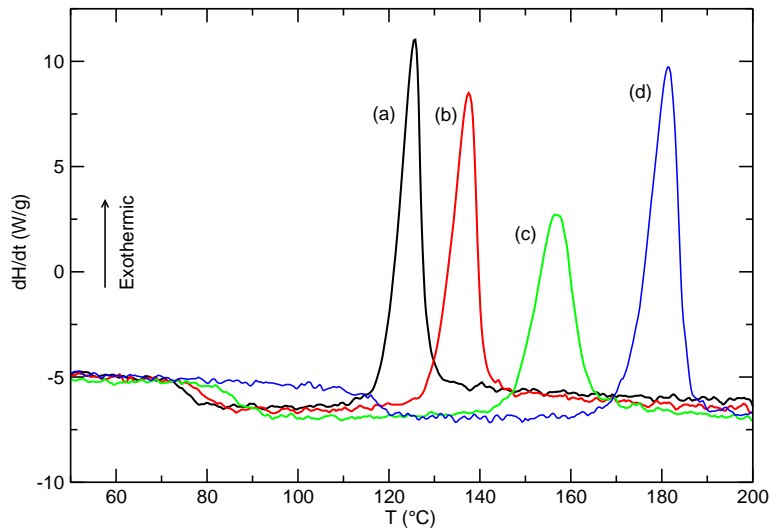}
\end{center}
\caption{DSC scan of as received material. Heating rate $2.5$~$^\circ$C/min: (a) PET, (b) 90/10 PET/PEN, (c) 70/30 PET/PEN, (d) PEN.}
\label{dsc-as_received}
\end{figure}

In figure~\ref{dsc-as_received} 
the DSC scans of the materials before being processed into sheets is presented. Approximate $T_g$ values can be deduced from this figure. This values are presented in table~\ref{tege}. 
\begin{table}
\caption{Approximate values of dynamic $T_g$ at $2$~$^\circ$C/min obtained from DSC.}
\label{tege}
\begin{center}
\begin{tabular}{|c|c|c|c|}\hline
PET & 90/10 & 70/30 & PEN \\ \hline
$76$ & $80$ & $87$ & $118$ \\ \hline
\end{tabular}
\end{center}
\end{table}
A single glass transition is observed on all the plots. This indicates that there has been enough polymerization to compatibilize the blends as expected from $^1$H--NMR results.

\subsection{TSDC--RMA results}

\begin{figure}
\begin{center}
\includegraphics[width=10cm,clip]{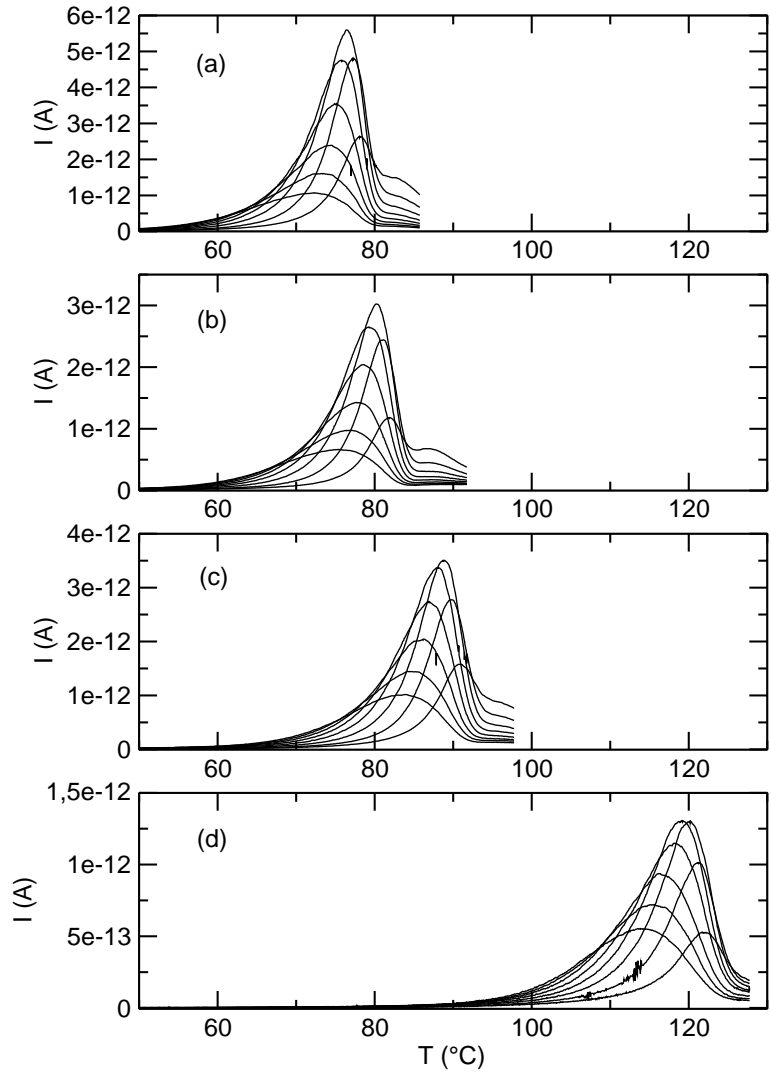}
\end{center}
\caption{Plot of the TSDC spectra obtained using NIW: (a) PET, $T_p$ from $78$ to $64$~$^\circ$C in 2~$^\circ$C steps, (b) 90/10 PET/PEN, $T_p$ from $82$ to $68$~$^\circ$C in 2~$^\circ$C steps, (c) 70/30 PET/PEN, $T_p$ from $90$ to $76$~$^\circ$C in 2~$^\circ$C steps, (d) PEN, $T_p$ from $122$ to $108$~$^\circ$C in 2~$^\circ$C steps.}
\label{rma-niw}
\end{figure}

Figure~\ref{rma-niw} 
shows the depolarization currents obtained by TSDC/NIW measurements that make up a RMA. In all cases, we obtain almost elementary spectra, this is, spectra that records mainly the response of a single mode. These spectra represent the modes of the relaxation. We will call optimal poling temperature ($T_{po}$) to the poling temperature employed to obtain the spectrum that yields the maximum current value of all the spectra, and temperature of the maximum ($T_{max}$) to the temperature at which that maximum current takes place \cite{peta}. 

The first observation is that for each material $T_{max}$ follows the same pattern as $T_g$ in figure~\ref{dsc-as_received}. In fact, it is known that $T_{max}$ corresponds to the value of $T_g$ at the rate of the heating ramp \cite{peta}. Figure~\ref{rma-niw} only shows one glass transition temperature in each sample, as it was observed by DSC. This fact supports the idoneity of the copolymerization--blending process carried out. In figures~\ref{rma-niw}(a), \ref{rma-niw}(b) and \ref{rma-niw}(c) (samples with some PET content), the presence of a small $\rho$ peak can be seen at about $8$~$^\circ$C above $T_{max}$.

Besides these observations, another interesting point of these scans is the evolution of the shape of the peaks. As PEN content increases in the samples it can be noticed a slight increase in the symmetry of the current peaks obtained using lower poling temperatures (high frequency modes). More symmetrical peaks indicate less cooperativity in the relaxation, as we will discuss further ahead. 

\subsection{Modeling of TSDC--RMA}

Modeling of TSDC data have been performed assuming that, if the thermal window is narrow enough, the response will be due almost entirely to just one elementary mode. As such, it will be well described by a single relaxation time ($\tau$). The intensity current that corresponds to an elementary mode $i$ of the $\alpha$ relaxation is given by~\cite{bucci}
\begin{equation}
I_i(t) \propto \exp \left[ - \int_{t0}^t \frac{dt'}{\tau_i(t')} - \ln[\tau_i(t)] \right].
\label{current}
\end{equation}
In this equation, the dielectric relaxation time is considered to be the same as the structural relaxation one. The parameters that correspond to each mode are obtained fitting a calculated depolarization current $I(T)$ obtained from equation~\ref{current} to experimental data. There are several models for $\tau$ that can be used to modelize the current.

In the Arrhenius model the relaxation time is assumed to obey the equation:
\begin{equation}
\tau(T)=\tau_0 \exp \left(\frac{E_a}{RT}\right),
\label{arrheniuslaw}
\end{equation}
where the activation energy $E_a$ and the pre--exponential factor $\tau_0$ have a simple physical interpretation~\cite{gert}. Some values of $E_a$ and $\tau_0$ obtained from the fits are reproduced in table~\ref{valors}. 
\begin{table}
\caption{Some representative parameters obtained from the mathematical fit of TSDC/WP curves to the Arrhenius model.}
\label{valors}
\begin{center}
\begin{tabular}{|c|c|c|c|c|c|} \hline
$\tau_0$ & $E_a$ & $E_a$          & $P$       & $T_p$ \\
(s)      & (eV)  &  (kJ/mol) & (C/m$^2$) & (K)   \\ \hline
\multicolumn{5}{|c|}{PET} \\ \hline
$0.189 \times 10^{-59}$ & $4.28$ & $413$ & $2.33 \times 10^{-4}$ & $351.15$ \\
$0.863 \times 10^{-42}$ & $3.05$ & $294$ & $3.20 \times 10^{-4}$ & $347.15$ \\
$0.362 \times 10^{-13}$ & $1.07$ & $103$ & $6.10 \times 10^{-5}$ & $333.15$ \\ \hline
\multicolumn{5}{|c|}{90/10} \\ \hline
$0.361 \times 10^{-61}$ & $4.44$ & $428$ & $1.17 \times 10^{-4}$ & $355.15$ \\
$0.400 \times 10^{-42}$ & $3.10$ & $299$ & $1.78 \times 10^{-4}$ & $351.15$ \\
$0.564 \times 10^{-14}$ & $1.13$ & $109$ & $3.98 \times 10^{-5}$ & $337.15$ \\ \hline
\multicolumn{5}{|c|}{70/30} \\ \hline
$0.122 \times 10^{-58}$ & $4.37$ & $422$ & $1.41 \times 10^{-4}$ & $363.15$ \\
$0.370 \times 10^{-44}$ & $3.32$ & $320$ & $2.23 \times 10^{-4}$ & $359.15$ \\
$0.185 \times 10^{-15}$ & $1.26$ & $122$ & $5.99 \times 10^{-5}$ & $345.15$ \\ \hline
\multicolumn{5}{|c|}{PEN} \\ \hline
$0.264 \times 10^{-49}$ & $4.03$ & $389$ & $6.60 \times 10^{-5}$ & $393.15$ \\
$0.115 \times 10^{-33}$ & $2.80$ & $270$ & $1.05 \times 10^{-4}$ & $387.15$ \\
$0.218 \times 10^{-16}$ & $1.43$ & $138$ & $2.93 \times 10^{-5}$ & $373.15$ \\ \hline
\end{tabular}
\end{center}
\end{table}
These values correspond to a low--frequency mode, the optimal polarization mode and a high--frequency mode. 

\begin{figure}
\begin{center}
\includegraphics[width=10cm,clip]{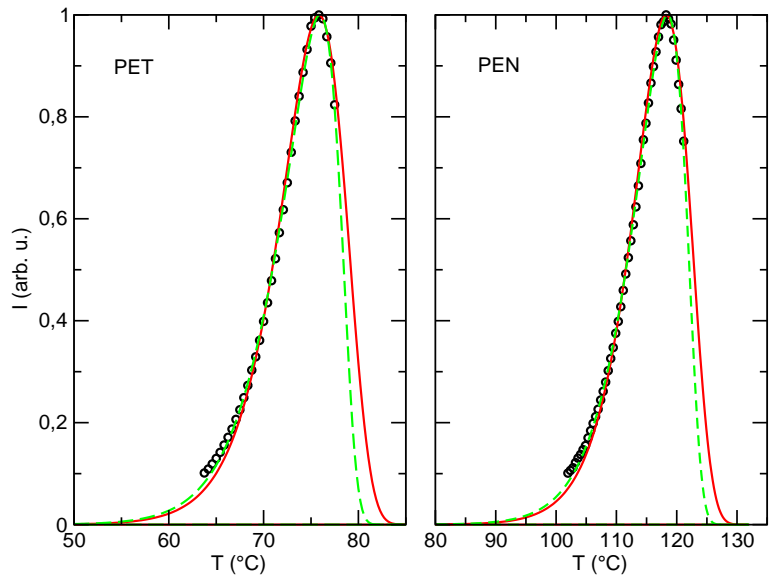}
\end{center}
\caption{Fits of the Arrhenius model (continuous line) and the TNM model (dashed line) to experimental data of the optimal polarization mode (circle) for PET and PEN.}
\label{ajust}
\end{figure}

The fit of the Arrhenius model for the optimal polarization mode of PET and PEN is shown in figure~\ref{ajust}.
The fit is pretty satisfactory in spite of the fact that the Arrhenius model is not the best model for a cooperative relaxation. Models such as Vogel--Tammann--Fulcher~\cite{vtf} or TNM are more commonly used to modelize the $\alpha$ relaxation. Nevertheless, both include the Arrhenius model as limit case and have three adjustable parameters, one more than Arrhenius, so we can use Arrhenius as a convenient approximation to get the big picture and study later the aspects that go beyond this model with a more capable model like TNM.

The existence of multiple modes is modelized through a distribution of relaxation times (DRT). This DRT can be described in terms of a few parameters \cite{kww} or, more accurately, we can assign a weight to each mode that quantifies the contribution of that mode \cite{petrma}. Within a RMA we can identify the weight of a mode with the product of the polarization of the sample by the polarization temperature \cite{laca}.

In spite of the fact that the Arrhenius model describes the $\alpha$ relaxation in a simple way (and with a considerable ease of interpretation), well known non--linear phenomena like thermal history effects and physical aging can not be explained by this model. The simplest extension to the Arrhenius model that incorporates this parameter is the Tool--Narayanaswamy--Moynihan (TNM) model~\cite{moynihan}.
\begin{equation}
\tau(T, T_f) = \tau_0 \exp \left(\frac{x E_a}{RT}\right) \exp \left[\frac{(1-x) E_a}{RT_f}\right].
\label{tnm}
\end{equation}
In this model the separation between $T$ and $T_f$ is introduced through a non--linearity parameter $x$ ($0 \leq x \leq 1$). The other two parameters are the pre--exponential factor $\tau_0$ and the activation energy $E_a$. The evolution of $T_f$ with time and temperature is introduced assuming an ideal--viscous return to equilibrium that depends on the relaxation time $\tau(t)$ by the equation
\begin{equation}
\frac{dT_f}{dt} = \frac{1}{\tau(t)}(T - T_f).
\label{viscosity1}
\end{equation}
Equations~\ref{tnm} and~\ref{viscosity1} are coupled. As a consequence, numerical methods are required to integrate them \cite{petrma}. Some values of $E_a$, $\tau_0$ and $x$ obtained during the fitting processes are reproduced in table~\ref{valors2}. 
\begin{table}
\caption{Some representative parameters obtained from the mathematical fit of TSDC/WP curves to the TNM model.}
\label{valors2}
\begin{center}
\begin{tabular}{|c|c|c|c|c|c|} \hline
$\tau_0$  & $E_a$  & $E_a$         & x & $P$       & $T_p$ \\ 
(s)       & (eV)   & (kJ/mol) &   & (C/m$^2$) & (K)   \\ \hline
\multicolumn{6}{|c|}{PET} \\ \hline
$0.259 \times 10^{-81}$ & $5.78$ & $558$ & $0.70$ & $2.15 \times 10^{-4}$ & $351.15$ \\
$0.702 \times 10^{-57}$ & $4.07$ & $393$ & $0.73$ & $2.94 \times 10^{-4}$ & $347.15$ \\
$0.362 \times 10^{-13}$ & $1.07$ & $103$ & $1.00$ & $6.10 \times 10^{-5}$ & $333.15$ \\ \hline
\multicolumn{6}{|c|}{90/10} \\ \hline
$0.278 \times 10^{-84}$ & $6.05$ & $584$ & $0.69$ & $1.07 \times 10^{-4}$ & $355.15$ \\
$0.700 \times 10^{-55}$ & $3.98$ & $384$ & $0.76$ & $1.67 \times 10^{-4}$ & $351.15$ \\
$0.564 \times 10^{-14}$ & $1.13$ & $109$ & $1.00$ & $3.98 \times 10^{-5}$ & $337.15$ \\ \hline
\multicolumn{6}{|c|}{70/30} \\ \hline
$0.271 \times 10^{-77}$ & $5.70$ & $550$ & $0.74$ & $1.33 \times 10^{-4}$ & $363.15$ \\
$0.137 \times 10^{-59}$ & $4.41$ & $425$ & $0.72$ & $2.07 \times 10^{-4}$ & $359.15$ \\
$0.186 \times 10^{-15}$ & $1.26$ & $122$ & $1.00$ & $5.99 \times 10^{-5}$ & $345.15$ \\ \hline
\multicolumn{6}{|c|}{PEN} \\ \hline
$0.575 \times 10^{-71}$ & $5.70$ & $550$ & $0.63$ & $5.95 \times 10^{-5}$ & $393.15$ \\
$0.220 \times 10^{-44}$ & $3.61$ & $348$ & $0.75$ & $9.88 \times 10^{-5}$ & $387.15$ \\
$0.219 \times 10^{-16}$ & $1.43$ & $138$ & $1.00$ & $2.93 \times 10^{-5}$ & $373.15$ \\ \hline
\end{tabular}
\end{center}
\end{table}
The selected values follow the same criteria as those of table~\ref{valors}. The fit of the TNM model for the optimal polarization mode of PET and PEN is also shown in figure~\ref{ajust}. There is a slight but noticeable improvement over the fit with the Arrhenius model. It can be seen that the TNM model deals better with the asymmetry of the current peak, not surprisingly since it was conceived with the structural relaxation in mind.

\subsection{Discussion of the TSDC--RMA modeling outcome}

\begin{figure}
\begin{center}
\includegraphics[width=10cm,clip]{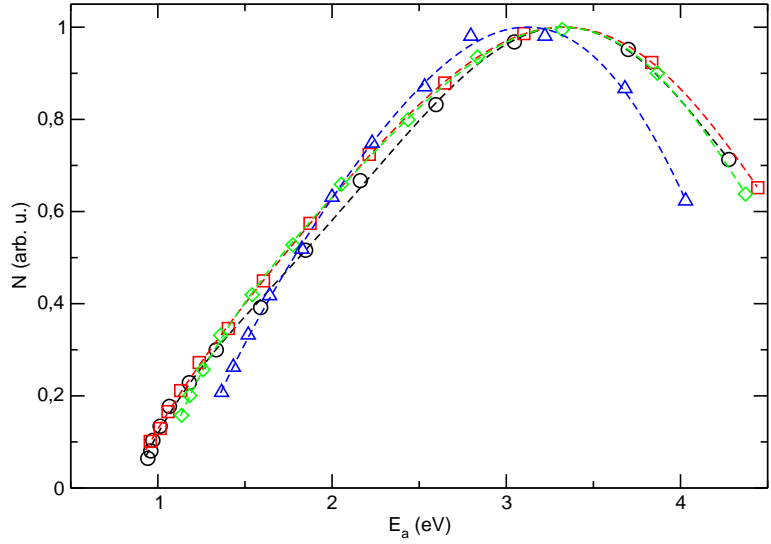}
\end{center}
\caption{Weight of each mode against its activation energy as obtained from the NIW RMA fits to the Arrhenius model: (circle) PET, (square) 90/10 PET/PEN, (diamond) 70/30 PET/PEN, (triangle) PEN. The weights of each material are normalized so the estimated maximum value is $1$.}
\label{weights-arr}
\end{figure}

We will begin discussing the results of the Arrhenius model. In figure~\ref{weights-arr} 
it can be seen the weight of each mode $N$ against its activation energy $E_a$. The weights are normalized so the estimated maximum weight is equal to~$1$. A fifth--order polynomial regression is used to estimate the shape of the distribution. We can hardly see any change in these curves between the samples. 

The activation energy that corresponds to the mode with maximum weight is around $3.3$~eV in all the cases except for the PEN sample that shows a slightly lower value ($3.1$~eV). The high temperature side of the curve also differs in the PEN sample from the other ones, although this fact is probably due to the presence of the $\rho$ peak in the samples with PET, that may influence the calculations. The shape of the distribution is noticeably asymmetric, with a steeper decrease on the low--frequency side. The only remarkable differences between materials appear in the lowest $E_a$ value obtained in each curve that show a progressive increase as PEN content does, from $1.0$ to $1.4$~eV.

\begin{figure}
\begin{center}
\includegraphics[width=10cm,clip]{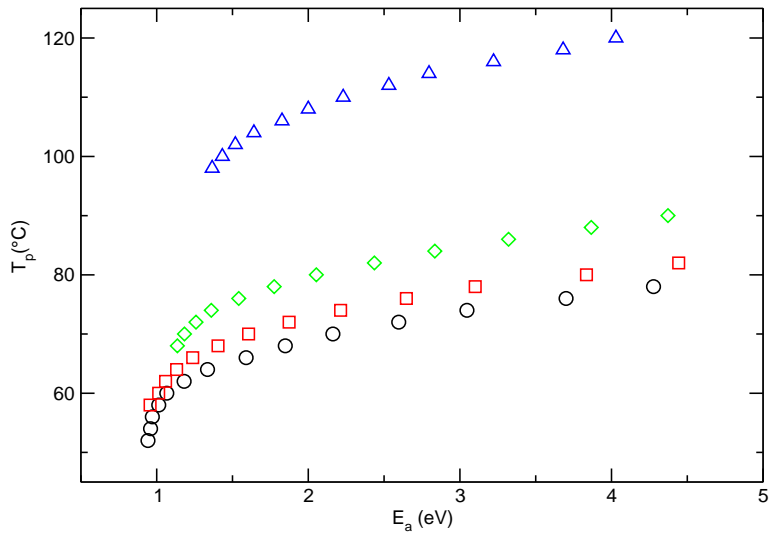}
\end{center}
\caption{Poling temperature at which each mode is obtained against its activation energy as obtained from the NIW RMA fits to the Arrhenius model: (circle) PET, (square) 90/10 PET/PEN, (diamond) 70/30 PET/PEN, (triangle) PEN.}
\label{cutoff-arr}
\end{figure}

This can be seen better in figure~\ref{cutoff-arr}, 
where the poling temperature at which each mode is obtained is plotted against its activation energy. The low--frequency part of the relaxation (higher $E_a$) is difficult to be study because there are too few points due to the asymmetry of the relaxation. Instead, the high--frequency part of the relaxation (lower $E_a$) has plenty of points that allow to see that there is a cutoff activation energy. This fact means that there are  no chain segments activated below certain value of $E_a$. This lower value of the activation energy increases with PEN content, from $1.0$ to $1.4$~eV as stated above.

So which is the reason why the relaxation time becomes larger and the glass transition shifts to higher temperatures as the PEN content increases? A plot of $E_a$ in front of $\tau_0$ is called a compensation plot and usually gives a straight line when plotted on a log--linear scale. The cause of this relationship is unclear. Some authors think that it is characteristic of cooperative relaxations \cite{consider} while others point out that it will appear whenever there is a sharp increase in the activation energy in the DRT \cite{sharp}. Nevertheless we are going to use the compensation plot just to compare the behavior of the relaxation time of the materials. Compensation behavior in polyesters has already been reported \cite{copos3}.

\begin{figure}
\begin{center}
\includegraphics[width=10cm,clip]{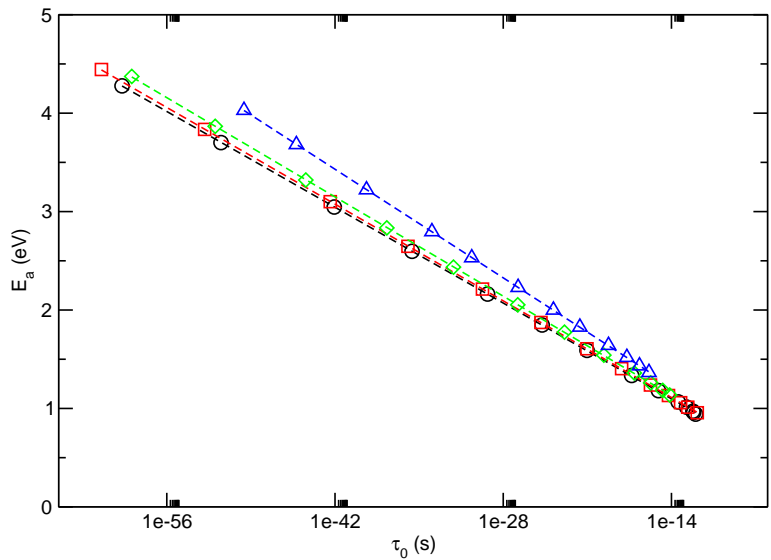}
\end{center}
\caption{Activation energy of each mode against its pre--exponential factor as obtained from the NIW RMA fits to the Arrhenius model: (circle) PET, (square) 90/10 PET/PEN, (diamond) 70/30 PET/PEN, (triangle) PEN.}
\label{compensation-arr}
\end{figure}

In the compensation plot presented in figure~\ref{compensation-arr} 
we can see that each polymer has its own compensation slope. The change in this slope is what enlarges the relaxation time because it gives a larger pre-exponential factor if activation energies are the same. Therefore, the change in the relaxation time is mainly due to a change in the pre-exponential factors, not in the activation energies.

A possible interpretation of these results can be the following one. We can assume that the activation energy of the relaxation, $E_a$, is mainly associated with the energy necessary to break hydrogen bonds between close chains, and/or to regain flexibility in bonds of the main polymer chain. In the framework of this assumption, all the materials (PET, PEN and their copolymers) would show similar values of $E_a$ due to the high similarity in their chemical structure. Once some segments of the polymer chain regain mobility in this way, the increased complexity of PEN monomers will cause the chain segments to move slower resulting in an increased relaxation time, as our results show.

Also, figure~\ref{compensation-arr} hints the existence of a point where the compensation plots of each material could join or, at least, come very close. This would mean that whatever mechanism is responsible for the increase of the pre-exponential factor is more effective at low frequencies (high activation energies). 

\begin{figure}
\begin{center}
\includegraphics[width=10cm,clip]{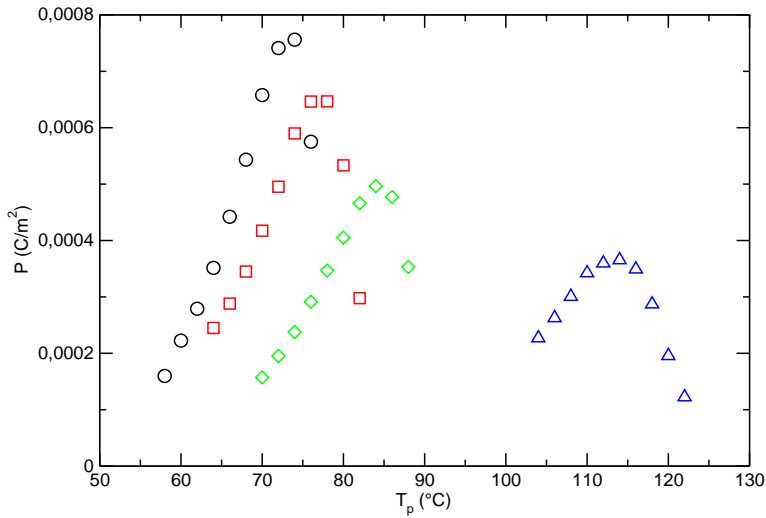}
\end{center}
\caption{Polarization of each mode against its poling temperature as obtained from the WP RMA: (circle) PET, (square) 90/10 PET/PEN, (diamond) 70/30 PET/PEN, (triangle) PEN.}
\label{polar-wp}
\end{figure}

Since saturation can not be attained using the NIW poling scheme, an additional RMA was performed by WP with $5$~min polarization time to obtain data about polarizability (see section~\ref{exp} for details). A correction factor has been applied to take into accounts deviations of the samples thickness from the mean $0.5$~mm value. Spectra obtained in these experiments (not plotted) almost exactly resemble the ones obtained using NIW poling scheme, although in this case delivering twice as much depolarization current as in the previous case. Polarization against poling temperature is presented in figure~\ref{polar-wp} 
for the four materials. A consistent trend is found compatible with a lower polarizability due to an increased stiffness when PEN content increases.

In spite of the interest of these results, the Arrhenius model does not take into account physical aging since the relaxation time has no dependence on the history of the sample. To quantify this effects the TNM model introduces a non--linearity parameter, $x$,  that quantifies the cooperativity of the modes. 

\begin{figure}
\begin{center}
\includegraphics[width=10cm,clip]{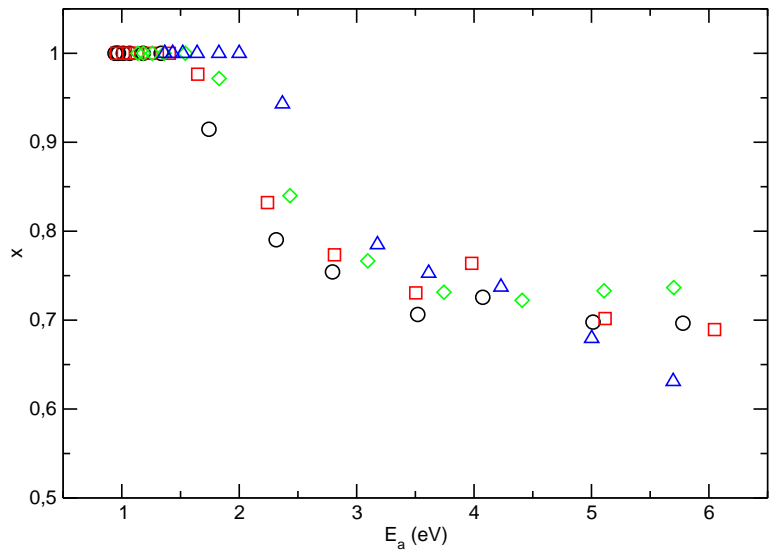}
\end{center}
\caption{Non--linearity parameter for each mode against its activation energy as obtained from the NIW RMA fits to the TNM model: (circle) PET, (square) 90/10 PET/PEN, (diamond) 70/30 PET/PEN, (triangle) PEN.}
\label{cooperativity-niw}
\end{figure}

The analysis of the non--linearity parameter has to cope with some uncertainty in the fits but it reaches similar values (around $x=0.7$) in all the samples, as it can be seen in figure~\ref{cooperativity-niw}. 
The modes activated around the optimum polarization temperature and above ($E_a$ of 3~eV and above) are the most cooperative. As we approach the high--frequency part of the relaxation (lower $E_a$) the cooperativity shrinks until it becomes arrhenius--like in all the cases. This happens earlier as PEN content increases, maybe due to the bulky naphtalene ring of the PEN chain, that makes more difficult that long chain segments move in a cooperative way. The analysis of the other parameters coincides with the Arrhenius results and is not discussed again.

\begin{figure}
\begin{center}
\includegraphics[width=10cm,clip]{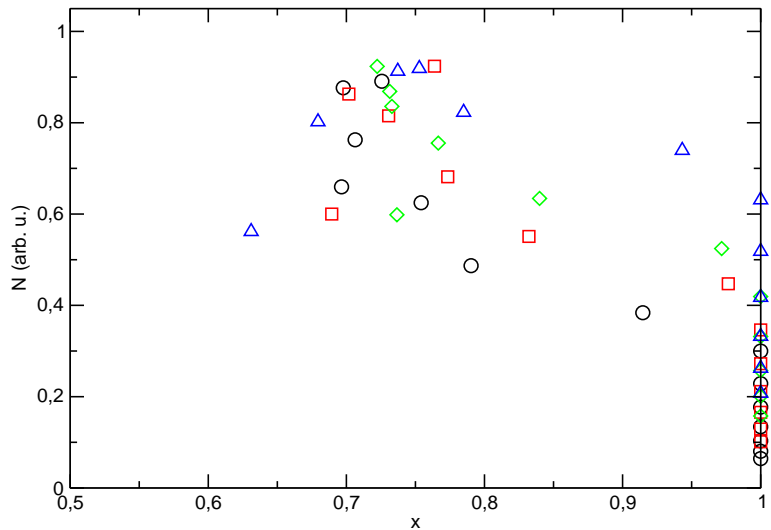}
\end{center}
\caption{Weight of each mode against its non--linearity parameter as obtained from the NIW RMA fits to the TNM model: (circle) PET, (square) 90/10 PET/PEN, (diamond) 70/30 PET/PEN, (triangle) PEN.}
\label{nx-niw}
\end{figure}

We can gain further insight on this matter if we represent the weight of each mode in terms of the value of its non--linearity parameter, as seen in figure~\ref{nx-niw}. 
It can be seen that in PEN the weight of the non-cooperative modes is almost as large as the weight of the cooperative ones. In spite of this fact the non--linearity of the most important modes is similar in all the materials with a value around $0.7$.

\subsection{Simulation of DSC curves}

Compared with dielectric loss spectroscopy, TSDC data is well suited to be compared with calorimetric data. The depolarization current is recorded when the sample is recovering structural equilibrium, just as it happens in DSC. Instead dielectric loss experiments with an equivalent frequency, around $10^{-6}$~Hz \cite{equivalent}, take too much time to be useful to study the dynamics of the structural relaxation. Moreover, the thermal history of a TSDC experiment and a DSC history are very similar if not the same. This means that given a complete enough model the normalized calorific capacity around the glass transition can be reproduced from the obtained parameters. TNM is fine for this purpose since it takes into account the structural state of the system through the fictive temperature \cite{petrma} and therefore it can reproduce memory effects or physical aging.

The normalized calorific capacity is obtained from the fictive temperature~\cite{cpnorm} through
\begin{equation}
c_p^n = \frac{dT_f}{dT}.
\end{equation}
Within the TNM model, the fictive temperature is calculated at the same time as the relaxation time since equations~\ref{tnm} and~\ref{viscosity1} are coupled. In this way, a normalized calorific capacity is obtained for each mode. The normalized calorific capacity of the system can be assumed to be the weighted average of the normalized calorific capacity of each mode, with acceptable results~\cite{petrma}. Again, the weights are the product of the polarization of the mode and the polarization temperature~\cite{laca}.

\begin{figure}
\begin{center}
\includegraphics[width=10cm,clip]{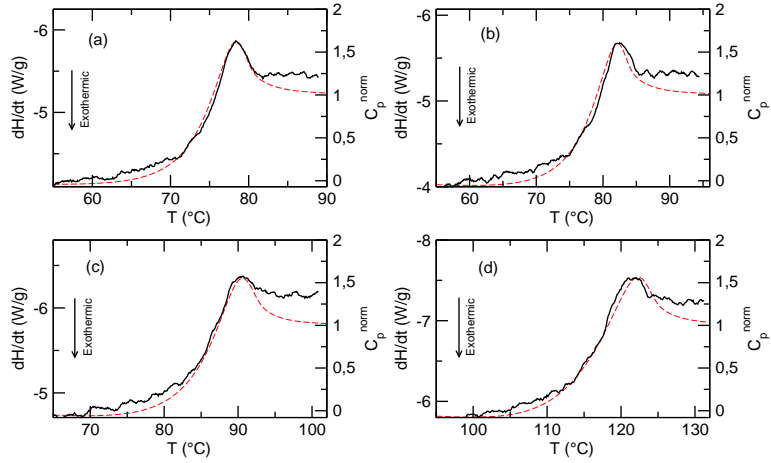}
\end{center}
\caption{DSC scans (continuous line, left scale) compared to predicted calorific capacity (dashed line, right scale): (a) PET, (b) 90/10 PET/PEN, (c) 70/30 PET/PEN, (d) PEN. The two vertical scales have been adjusted in order to ease the comparison.}
\label{dsc-niw}
\end{figure}

Figure~\ref{dsc-niw} 
presents a comparison between DSC data and a calculation of the normalized calorific capacity using the parameters obtained from TNM fits, for the four materials studied. The scale and the origin of the Y axis for experimental data have been chosen to allow a better comparison with the normalized data provided by the calculations. For the same reason, the direction of the exothermic behavior has been taken as downward and only the part close to the glass transition has been plotted. As it can be seen in the figure, there is agreement on the glass transition region.

\begin{figure}
\begin{center}
\includegraphics[width=10cm,clip]{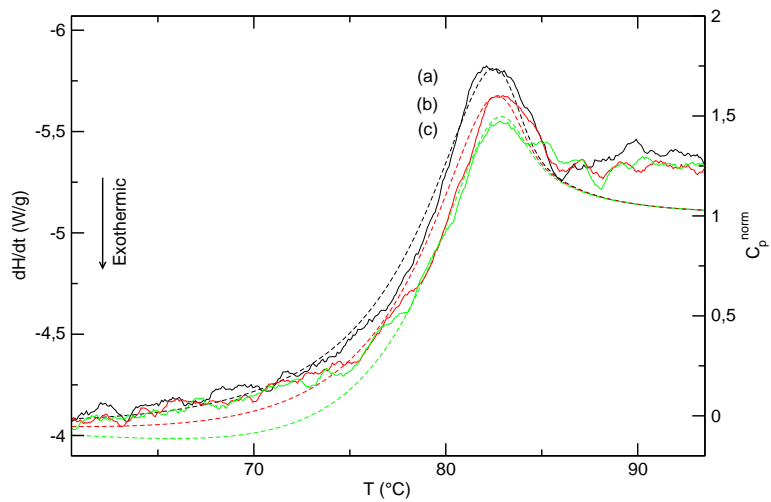}
\end{center}
\caption{DSC scans (continuous line, left scale) compared to predicted calorific capacity (dashed line, right scale) for PET with different cooling rates: (a) $1.25$~$^\circ$C/min, (b) $2.5$~$^\circ$C/min, (c) $5$~$^\circ$C/min. The two vertical scales have been adjusted in order to ease the comparison.}
\label{dsc-refr}
\end{figure}

The most striking feature of this approach is the ability to reproduce memory effects, as demonstrated in figure~\ref{dsc-refr}. 
In this figure, the results of the calculation of the normalized calorific capacity are compared with DSC data for several cooling rates but the same heating rate. The thermal history is the same in the TSDC experiments with the NIW poling scheme described in table~\ref{parameters-niw} except for the rate of the cooling ramp that has three different values ($1.25$~$^\circ$C/min, $2.5$~$^\circ$C/min, and $5$~$^\circ$C/min). The trends given by the varying cooling rate are well--reproduced giving us a demonstration of the capabilities of the TNM model. Lets recall that the Arrhenius model alone is not able to cope with memory effects since the relaxation time depends only on the temperature of the sample at a given time and therefore it would give the same results whenever the heating ramp is the same.

\section{Conclusions}

From the characterization of the blends follows that enough transesterification took place during the melt processing to compatibilize the PET/PEN blends. As a consequence these materials present a single glass transition and in TSDC spectra there is a single peak that corresponds to the $\alpha$ relaxation.

$T_{max}$, the temperature at which the maximum intensity appears when the sample is poled at $T_{po}$, follows the same pattern as $T_g$ obtained from calorimetric scans. This is not surprising, since the $\alpha$ relaxation is the dielectric counterpart of the glass transition.

The RMA procedure allows us to analyze the $\alpha$ relaxation as elementary modes. Each depolarization current (spectrum) is mainly the response of an elementary mode that can be fitted to obtain the activation energy and the pre--exponential factor of that mode. Within the TNM model also a non--linearity parameter is obtained.

Activation energies almost do not change, although there is a slight and non--significant decrease in the $E_a$ of the optimally poled mode for PEN with regards the other materials. The change in the relaxation time is due mainly to changes in the pre--exponential factor. The overall shape of the distribution of relaxation times is asymmetric with a much steeper decrease on the low--frequency side.

The compensation law is fulfilled in all cases with a characteristic slope that increases with the content of PEN. An interpretation has been given based on the energy needed to activate the process and the frequency at which it takes place. In any case, for high frequencies the compensation plots of the four materials tend to join.

There is an activation energy cutoff on the high frequency side. The energy of this cutoff increases with the content of PEN. On the other hand, the low frequency side is difficult to study because it overlaps with the $\rho$ relaxation and also because of the asymmetry of the distribution of relaxation times.

Polarizability decreases with larger PEN content as expected because of the increased stiffness of the polymer backbone.

The non--linearity parameter reaches values around $0.7$ for modes obtained poling at temperatures close to the $T_{po}$. This is another value, together with activation energy, that does not seem to differ between the four materials studied. Instead, the way the cooperativity changes as we consider modes with a higher frequency depends on the PEN content. High frequency modes are always Arrhenius--like ($x=1$). As the content of PEN increases the weight of these modes and also of the modes with an intermediate $x$ is larger in comparison with the weight of the central modes of the distribution. This is particularly visible in pure PEN. 

It should be noted that this analysis is possible thanks to the capability of TSDC to excite mainly one elementary mode when using a proper poling scheme. Unlike in approaches based on dielectric loss spectroscopy, we do not have to make any assumption on the shape of the distribution of relaxation times or use any kind of deconvolution. 

Another plus is that TSDC experiments can be made with the same thermal history as DSC experiments and therefore the return to structural equilibrium that takes place can be compared using an appropriate model. This allows us to reproduce calorimetric data, even memory effects, so we can rely on dielectric results as a proper description of the glass transition of PET/PEN blends.

\bigskip

\textbf{Acknowledgements} This work has been partially supported by projects PPQ2001-2764-C03-03 (CICYT) and 2009 SGR 01168 (AGAUR).

\newpage

\pagestyle{empty}\clearpage\clearpage\clearpage\clearpage\clearpage
\end{document}